# A Network Approach to the French System of Legal codes – Part I: Analysis of a Dense Network

Romain BOULET* · Pierre MAZZEGA · Danièle BOURCIER

*Romain BOULET (*to whom correspondence should be addressed*)
UMR ESPACE-DEV, IRD, 500 Rue Jean-François Breton, 34000 Montpellier, FRANCE
E-mail: romain.boulet@ird.fr

Pierre MAZZEGA
a) Laboratoire Mixte International Observatoire des Changements Environnementaux, UnB / IRD, Universidade de Brasilia, BRASIL, E-mail: pmazzega@gmail.com
b) Université de Toulouse; UPS (OMP), CNRS, IRD, Geosciences Environnement Toulouse, 14 Avenue E. Belin, 31400 Toulouse, FRANCE

Danièle BOURCIER
CERSA CNRS, Université de Paris 2, 10 rue Thénard, 75005 Paris, FRANCE
E-mail: daniele.bourcier@cersa.cnrs.fr

Abstract

We explore one aspect of the structure of a codified legal system at the national level using a new type of representation to understand the strong or weak dependencies between the various fields of law. In Part I of this study, we analyze the graph associated with the network in which each French legal code is a vertex and an edge is produced between two vertices when a code cites another code at least one time. We show that this network distinguishes from many other real networks from a high density, giving it a particular structure that we call *concentrated world* and that differentiates a national legal system (as considered with a resolution at the code level) from *small-world* graphs identified in many social networks. Our analysis then shows that a few communities (groups of highly wired vertices) of codes covering large domains of regulation are structuring the whole system. Indeed we mainly find a central group of influent codes, a group of codes related to social issues and a group of codes dealing with territories and natural resources. The study of this codified legal system is also of interest in the field of the analysis of real networks. In particular we examine the impact of the high density on the structural characteristics of the graph and on the ways communities are searched for. Finally we provide an original visualization of this graph on an hemicyle-like plot, this representation being based on a statistical reduction of dissimilarity measures between vertices.

In Part II (a following paper) we show how the consideration of the weights attributed to each edge in the network in proportion to the number of citations between two vertices (codes) allows deepening the analysis of the French legal system.

# 1. Introduction

Generation after generation, the many national legal systems as well as the federal, Community or International legal systems are built by adding, revising, removing more or less important legal texts that cite each other and interfere under the requirement of legal coherence. A long and collective work, still continued, responding to the changes in the human societies it regulates, encourages or solicits, and to which many craftsmen contribute but none can claim to be the architect who has a complete and accurate picture of the produced building. The mathematics and computer sciences can contribute greatly to draw an overall picture of a legal system, promoting the development of new representations and understanding of law, less local or fractioned into specialty areas, more sensitive to its dynamic and evolution.

Codification has a long tradition in French Law. Historically, the Civil Code (1804) has inspired several continental law systems. Methodologically, the underlying rationale for codification (the grouping of texts of successive and dispersed articles in a single structure at a given time) has become a reference for legislators who want to re-codify or codify existing law. The codification process may include various forms more or less structured. In 1989 France has systematized a codification method characterized as being on the basis of established law and has created a commission to implement it. The European Union has taken some aspects of the method of French codification. Codification is also intimately related to the long standing problem of the control of legal complexity. Two reports were made by the Council of State (Conseil d'Etat 1991; Conseil d'Etat 2006) which participates in *a priori* control of writing texts, lambasting this complexity.

In an *a priori* control of drafting and writing of legal documents, the Constitutional Council has even considered that the unreasonable complexity of laws could make them unconstitutional: in its decision of 29 December on the Finance Bill for 2006, the Constitutional Council annulled outright Article 78 of the cap of tax loopholes on the grounds that the article would have "*reached a level of complexity that [it] infringes Article 14 of the Declaration of Human Rights and of the Citizen, which states that all citizens have the right to see [...] the necessity for public contribution*". The codification was seen as a way to reduce the complexity of the law by organizing more effective links between the provisions.

## 1.1 Analyzing the complexity of the law

Whatever their nationality, lawyers and legal practitioners now recognize the complexity of the law. Legal complexity raises substantive issues about the essence of norms (normativeness), about their relation to the legal facts and the "state of affairs" (Epstein, 1995) or as to the cognitive dimension of any norm which degree of complexity induces a diversity of behavior and therefore costs to ensure their mastery or use (Kaplow 1995). Tullock (1995) seeks to introduce the difficult problem of the accuracy that the legal norm should develop, a trade-off to be found between genericity, applicability and legal security.

Now a large amount of literature is devoted to the analysis of legal complexity, favoring different point of view or relying on a range of methodological approaches like the economic analysis of law, the use of concepts of the computational complexity theory, doctrinal analysis, etc. (see e. g. Schuck 1992, Kades 1997, Ruhl 1997, or more recently Pagallo 2006, Doat et al. 2007). The French State Council has underlined that legal complexity is opposed to several principles, some of them with a constitutional value, such as the legal security or accessibility to all citizens (Conseil d'Etat, 2006). However, an objective definition of the complexity of law is lacking and hence methods for its characterization and monitoring. The sciences that study complex systems, whether natural or artificial, provide concepts and tools that may be used to promote the emergence of new approaches to law and legal systems.

Today the approach of the legal complexity by graph does not pretend to tackle with all the relevant issues of legal complexity. It develops a rather minimalist strategy: focus first on identifying treatable questions (showing a degree and a type of complexity that can be formalized) and develop adequate methodologies and tools for its characterization (statistical "*measures*", Halmos, 1974) and analysis. In this way this strategy begins gradually to meet the needs expressed by many authors for decades (see e.g. Long and Swingen, 1987 that proposed some qualitative measurement of the complexity of tax law). The temporal evolution of legal systems via a network-based approach, discussed in Bommarito et al 2010b, Fowler et al., 2007 or Leicht et al., 2007, is also relevant to better understand this complex organization.

In this paper we shall focus on the law complexity as manifested by the numerous cross-citations between legal texts (see Bourcier and Mazzega, 2007a,b). Fowler et al. (2007) show the contribution of a network approach to identify the most important precedents of the Supreme Court of the United States that therefore inspired the legal culture of this country, their importance being measured in terms of centrality in a network of citations in a large set of majority opinions (rather than by counting the number of citations; see also Fowler and Jeon, 2088). Bommarito and Katz (2009, 2010a) analyze the United States Code as a large network where the links are citation links (for instance the Tax Code – Title 26 of the US Code – is linked to the Criminal Code – Title 18 – through more than two dozen of citations).

The French Department of Justice has funded research on codification (Bourcier, 1998). This research has attempted to systematize the contributions of technology to the codification and to assess the effects of computerization on the access to law. During the bicentennial of the Civil Code, a review of work on computers and codification has been reassessed (Bourcier, 2004). These studies have focused on analyzing the relationship between writing code and the organization of law. But there are no studies using these results to analyze the relationship between branches of law as captured by codes or by groups of codes. This work is undertaken and could contribute to the partial renewal of Legal Studies. We propose here to study an aspect of the relationship between branches of law. We do not go into details about the organization of a single code or about the structuring of its network of citations. We develop a more comprehensive global view of the French legal system. To do this we simply consider each code as a legal entity and we identify the citations between these codes. In the representation that we draw, each code is a vertex and in code X a reference to code Y induces an edge between the corresponding vertices of the graph.

However we did not keep all the codes that are published on the public site of French law LEGIFRANCE (Legifrance, 2009). The codes that were not produced following the procedure of legislative drafting (Secrétariat Général du Gouvernement, 2007) prepared by the Higher Commission for Codification currently in force were excluded from our analysis. For example the general tax code that is produced by another authority (the Ministry of Budget) following a specific procedure is not part of the 52 codes that we selected (see Table 1). On the other side there are codes that have not been affected by the 1989 codification policy, as the Civil Code, fundamental matrix of the right (which old fashioned numbering system is maintained), because they were an inspiration to define good coding practices. The analysis of this network aims at constructing a meta-representation of the relations between the codes. This mapping of the work done by the coders since the launch of the codification policy in 1989 may be of great interest to the Higher Commission for Codification, which plans, manages and coordinates both the codification work and the maintenance of codes.

### 1.2 Analyzing a dense network

The large number (hundreds or thousands) of vertices present in the graphs built from real contexts requires to rely on metrological analyses and on the research of communities (group of vertices) to better appreciate the architecture of the network. Common features of these networks appear as for example the small-world structure found in various fields of application like peer-to-peer networks (Latapy and Magnien 2008), lexical network (Gaume 2004), networks between firm directors (Davis *et. al* 2003) or protein networks (del Sol *et al.*, 2004). The characteristic of a small-world are that the average distance between two nodes of the network is low (also known as tight global connectivity) and secondly that the probability of existence of an edge between two nodes is higher if these two nodes are linked to a third one (we also say that the network has a strong local connectivity) (Watts, 2003). These large interaction networks also share a property of low density (note that fundamental notions such as "density" are recalled in sec. 1.3), giving them well-marked structures in the form of communities (sub-dense parts separated by sparse sections) allowing a simpler representation and making their interpretation easier.

However, a graph with few vertices (less than one hundred) will not be necessarily easy to analyze provided it is dense, plenty of links that induce a specific complexity that is not found in large low-density graphs. Indeed, the complexity does not lie in the great number of vertices to consider, but in the large number of edges that prevents from the straightforward recovery of separate communities. While most studies of real networks address the analysis of low density graphs, we here analyze a network with a low number of vertices but much higher density.

The graph associated with the network of code citations of the French legal system is shown in Figure 1. It is formed of 52 vertices (the 52 codes retained for our analysis - see above) and 531 edges. The production of a representation favoring its interpretation is not a trivial task. In Figure 1 we use for this representation a spring-force algorithm that better fills the space of representation and limits crossing edges. It remains that the calculation of structure indices and the search for a more explicit representation is needed to better exhibit and understand the structure underlying the codified legal system. In Section 2 we analyze indices obtained from a metrological approach taking into account the density of the graph and introduce a new class of networks: the concentrated world. In Section 3 we will tackle with the difficulty in using the traditional tools in the search for communities when applied to dense graphs. We propose in Section 4 a method for the interpretation of the network using a representation based on a dissimilarity index applied to the vertices of the graph. The major features of the network associated to the French legal system are discussed in Section 5 and some conclusions and perspectives are given in the last Section 6.

### 1.3 Usual notions in graph theory

For the reader not acquainted with graph theory, we remind here the meanings of some terms used in this study. The usual notions of network analysis present in this paper can be found in reference books (such as Brandes and Erlebach, 2005).

- A *graph G(V,E)* is defined by a set of vertices *V* and a set of edge *E*, an edge linking two vertices.
- A *sub-graph* induced by a subset $W \subset V$ of the set of vertices of the graph *G* is the graph whose set of vertices is *W* and the set of edges is constituted by all the edges of *G* linking two elements of *W*.
- Two vertices are *neighbors* if they are linked by an edge.
- The *neighborhood* $\Gamma_v$ of a vertex *v*, is the set of neighbors of *v*.
- The *degree* of a vertex is the number of its neighbors, that is $|\Gamma_v|$.
- The *density* of a graph is the ratio between the number of edges that actually exists in the graph by the total number of possible edges of the graph, that is, $\frac{2m}{n(n-1)}$ where *n* is the number of vertices of the graph and *m* is the number of edges.
- A *path* between two vertices $v_1$ and $v_k$ is a sequence of vertices $(v_1, v_2, ..., v_k)$ where the vertices $v_j$ and $v_{j+1}$ (for $j<k$) are linked by an edge. The *length* of the path is $(k-1)$.

- A *shortest path* between two vertices is a path of minimal length between these vertices. Several shortest paths can exist between two given vertices (with the same length).
- The *average path length* is the mean of the length of the shortest paths between each pair of vertices (Watts and Strogatz, 1998).
- The *characteristic path length* of a graph is the median of the means of the shortest paths from a vertex to the other vertices (Watts and Strogatz, 1998).
- The *diameter* of a graph is the longest of the shortest paths between two any vertices of the graph.
- A *clique* of a graph $G$ is a complete sub-graph of $G$, that is, a sub-graph in which all the vertices are pair-wise linked by an edge. A clique with $k$ vertices is also called a $k$-clique. A clique is maximal if it is not contained into another clique.
- An *Erdös-Rényi random graph* with parameters $n \in N$ and $p \in [0,1]$, denoted by $G(n, p)$, is a graph with $n$ vertices where each edge exists with a uniform probability $p$ (Erdös and Rényi, 1959).
- The adjacency matrix of a graph $G$ with $n$ vertices labeled from $1$ to $n$ is the matrix whose $(i,j)^{th}$ entry is $1$ if there is an edge between the vertices $i$ and $j$ and $0$ otherwise.

## 2. Structural analysis

### 2.1 Constructing the FLC network

The codes citation network of the French legal system is constructed as follows. Each of the 52 codes adopted is interrogated in turn on the site of Legifrance (2009), considering only their legislative part. For each code we do an automatic search for the occurrence of the word "code". Each of these citations is edited so as to verify that this is a citation of one of 52 legal codes (and not e. g. the occurrence of the string "code", use of word "code" in a sense irrelevant to our analysis, or one of the codes produced without following the legistic criteria established by the High Commission of Codification) and to identify that or those cited codes. The number of citations between two codes is well recognized. As we explained earlier in this article we do not consider these numbers, but only the existence of at least one quotation (which we represent as an undirected and un-weighted link). The graph whose edges are weighted by the number of citations is analyzed in Part II of this study.

The representation of the graph associated with the system of French legal codes (FLC) given in Figure 1 is produced with a *spring-force* algorithm. We first notice that this system is forming only one connected component (plus the "code of the honor legion" sharing no citation with any of the other codes). The code of the monetary media (with label INM; see Tab.1) presents a single connection with the rest of the FLC system. But at first glance no obvious structure appears (on this particular representation of the graph but as well on many alternative representations that we can draw). A more detailed analysis of the graph is required in order to have an overview of the properties (structural measures) of the FLC system and to identify code communities (if any).

### 2.2 Structural measures

Several quantitative measures can be used to characterize the structure of a network. Among these measures we here mainly rely on:
- General indices like the number $n$ of vertices, the number $m$ of edges, the density $d$ (see Sec. 1.3) and the mean degree $k = 2m/n$.
- Small-world indices (Watts and Strogatz, 1998), that is:
  - indices measuring the global connectivity such as the diameter $D$, the mean of shortest paths $\bar{l}$ or the characteristic path length $L$ (Watts, 2003);
  - indices measuring the local connectivity such as the first clustering coefficient $C_1$ which is the mean of the densities of neighborhood of vertices and the second clustering coefficient $C_2$ which is the ratio between the number of triangles and the number of connected triples.
- Centralization measures: according to Freeman (1979) three centralizations measures of a network can be defined based on three notions of vertex centrality (or power):
  - the degree centrality proportional to the degree of a vertex;
  - the betweenness centrality proportional to the number of shortest paths going through a vertex;
  - the closeness centrality, inversely proportional to the mean of shortest paths beginning at a given vertex.

A centralization measure lies between 0 (regular and not centralized network) and 1 (highly centralized network). In order to appreciate how high or low these indices are we compare them to an Erdös-Rényi random graph and to other real networks. In Table 2 we summarize the estimates of these measures for the FLC system, for the associated Erdös-Rényi random graph (which is in fact the mean of indices obtained by generating 10000 random graphs) and for three networks already studied by various authors, say a network of collaborations between mathematicians (Grossman, 2002), a Peer-to-Peer network (Latapy and Magnien, 2008) and a medieval social network (Boulet, 2008).

Let us compare these measure estimates for the different graphs. With only 51 codes in the connected component, the FLC system is a small system (as compared for example with the peer-to-peer network P2P with several millions of vertices). But its density is high, with a value of 0.416, the other graphs presenting densities of at most a few percent (medieval social network). This distinguishing property is to be related to the mean degree found in the FLC system: in average, a code is connected (being cited or citing) with more than 20 codes ($k$=20.8). The P2P network has a high mean degree but a very low density because of the plethora of vertices.

This high density of the FLC system also impacts the other measure estimates: a) the very low value (1.595) of the mean of shortest paths (*i.e.* the fact that it is most likely to be able to go from one code to any other code in one or two references) can be seen as a consequence of the high density: among the large amount of edges present in the network, some of them are creating shortcuts for the linkage of pairs of vertices (the same holds for the value of the characteristic path length *L and the diameter*); b) the values of the clustering coefficients and centralization measures of the FLC network are systematically and significantly higher than those of the population of the equivalent random graphs but the clustering coefficients are also high as a result of the high connectivity of the FLC network, a global property that is almost perceptible on Fig. 1. c) the betweenness centrality is high in the FLC network because there are many shortest paths with length 1 or coexisting shortest paths between any two codes with length 2.

Small-world networks are graphs with a tight global connectivity (that is a low diameter or a low mean of shortest paths) and a high local connectivity (high clustering coefficients – see *e.g.* the network of mathematicians in Table 2). The FLC network seems to have these properties but it is implicitly admitted that a small-world network is a globally sparse network (ie with a low density), which contrasts with the FLC network. Moreover the values of clustering or shortest path length for the FLC network are influenced by the very high density of this network (which is a global value of the network) whereas the local properties of a small world network (clustering) cannot be inferred by the global property of the density. Therefore we do not qualify the FLC network as a small-world network.

However we expect that some hidden code communities are structuring the FLC system: the reason is that the clustering coefficient $C_1$ (mean density of the neighborhoods of vertices) is quite higher than the average density *d* though we notice that the values of these measures are the same in the case of the random graphs. Indeed in a random graph the local density ($C_1$) is the same as the global one (*d*) whereas in a real network (and especially in a small world network) the graph is globally sparse but locally dense (in particular due to the presence of communities).

## 3. Finding communities

The network of references in the FLC system can reveal some preferences of inter-citations within groups of codes. The existence of such communities of codes would express the privileged interdependence between different legal domains. These communities are not immediately visible in the FLC system so that to identify them we use several algorithms. Community detection is an essential stage in understanding the architecture of a network and it constitutes a very active field of research in graph theory and network analysis. Some recent surveys (Fortunato, 2010; Porter et al., 2009) are detailing some of the algorithms and methods used to find communities in a network.

### 3.1 The rich-club

Another structural characteristic of networks is the presence or the absence of a rich-club (Colizza et al, 2006). A rich-club occurs in a network if the nodes with highest degree (also called 'rich nodes') tend to be strongly interconnected. The rich-club, if it exists, can be seen as a whole community with a central and influent role.

The analysis of the function $\varphi(r)$ which gives the density of the graph induced by the first $r$ vertices of highest degree (that is the vertices with highest degree centrality) (Zhou and Mondragon, 2004) and whose layout is given in Figure 3 reveals the presence of a single and influent pole called *rich-club*: the richest vertices are highly interconnected, forming a cohesive and influential group.

The seven codes with highest degree centrality are all connected together and then three codes come to the eighth place. Each of these three codes has a degree equal to 35 and is linked to the seven previous codes, forming cliques of order 8. These ten codes of highest degree are listed in Table 3. They form a dense sub-graph (only one edge is missing: the link between the environmental code - label ENV - and the code of employment - label TRAV) and constitute the rich-club. Moreover the betweenness and closeness centralities are represented are represented in Figure 3 as a function of the degree. We can see that the ten codes with the highest degree also have the highest betweenness centrality and highest closeness centrality. This emphasizes the central position of this influential group. On this point the FLC network contrasts with other real networks as it is not uncommon in real networks to have vertices with a low degree but with a high betweenness centrality measure (Newman, 2005; Boulet et al, 2008). We also note in Figure 3 that the Criminal code presents a betweenness centrality particularly high and can be seen as a code (and therefore legal domain) central to the French legal system but also a hinge between several legal domains.

As explained in the previous sub-section, the FLC network looks like a small world with regard to some of its connectivity characteristics but the density of links is much higher than what has been currently observed by researchers analyzing many networks or graphs representing natural or artificial systems but this high concentration of links gives the network a particular structure, different from a small world structure. Moreover it has a central rich club of codes concentrating also the betweenness and closeness centrality. For these reasons we propose to call this kind of network a "*concentrated world*".

### 3.2 Other communities

Discarding the 10 codes belonging to the rich club, which can be seen as a dominating community, can we separate the FLC network in a few components (sub-graphs) that present a high degree of internal connectivity? Shall we find hidden clusters of codes in the complex FLC network (see Figure 1)? Is there some underlying legal logic inducing such partitioning? In order to answer these questions we use three different well-known algorithms of graph partitioning whose aim is to maximize modularity. The modularity M measures how good is a partition $(V_1, V_2,…, V_k)$ and is defined as:

$$M = \sum_{i=1}^{k} \left[ \frac{m_i}{m} - \left( \frac{\sum_{u \in V_i} d(u)}{2m} \right)^2 \right] \quad (1)$$

where $m_i$ is the number of edges in $V_i$ and $d(u)$ denotes the degree of the vertex $u$. It measures the difference between the number of edges in a cluster and the number of expected edges. The value of M is high when the trial partition separates well-marked communities (if any) that structure the analyzed network. Three well-known algorithms are considered:

- A fast-greedy algorithm for partitioning which finds communities in graphs via a direct optimization of the modularity of trial aggregations of vertices (Clauset et al, 2004; Newman, 2004). At each step of the process we choose the aggregation of communities which results in the greatest increase of the modularity;
- A spectral graph partitioning based on the normalized Laplacian (von Luxburg, 2007). The normalized Laplacian of a simple graph is **L=D**$^{-1/2}$**(D-A)D**$^{-1/2}$ where **A** is the adjacency matrix (see Section 1.3) and **D** = diag(**A1**). The vertices are mapped in $R^k$ via their coordinates given by the first $k$ eigenvectors of **L**. This allows performing a partitioning algorithm such as a $k$-means algorithm, see (Hartigan and Wong, 1979) for instance, to find communities. The number of clusters is chosen in order to maximize modularity;
- The walk-trap algorithm (Pons and Latapy, 2005) based on random walks; the underlying idea is that a random walk on a graph will be trapped into a community after a certain number of steps. The proposed algorithm in (Pons and Latapy, 2005) begins to compute a distance between vertices through random walks on the graph. This distance yields a distance between group of vertices (communities) and then a hierarchical clustering is performed by merging two communities if they are close.

We perform each algorithm on the graph obtained by removing vertices belonging to the rich-club. The graph associated with the FLC system exhibiting the rich-club and the partitions obtained with these three algorithms are presented in Figures 4, 5 and 6 respectively. These figures are equivalent to Figure 1 but they show the modular structures found in the FLC network.

The rich-club is represented by a central rectangle. The communities obtained by the partitioning algorithms are represented by diamonds and disks represent the Honour Legion code (LGA; isolated vertex) on the one hand and Monetary media code (INM; pendant vertex) on the other hand. A weighted edge between two symbols indicates the total number of links between these two groups of codes. The acronym of each code is transcribed in the symbol of the community to which it belongs. For example in Figure 4, the ASF code belongs to the community of 12 codes most strongly linked to the rich club.

At first glance, the two graphs obtained with the fast-greedy (Fig.4) and spectral (Fig.5 algorithms are quite similar, the third graph (walk-trap algorithm) exhibiting a larger number of smaller communities. Of course on these figures the particular representations chosen for these graphs tend to emphasize the similarities but the lists of codes belonging to similar communities of Fig.4 and 5 are not identical (except of course for the rich-club which is preserved by the analysis). In fact the results differ depending on the process chosen to establish the partitioning. In contrast the non-uniqueness of optimal partitioning justifies our strategy of using several algorithms (based on different partitioning approaches).

These partitioning highlight two stable groups that is sets of codes belonging to the same class for the three considered partitions. These two groups can be seen as two stable communities of respectively 12 codes and 11 codes. The first group is also the second largest class obtained using the fast-greedy partitioning algorithm (Clauset et al, 2004). The second stable community also coincides with the second largest class in the partition obtained by the method of random walks (Pons and Latapy, 2005). An interpretation of these two communities is outlined in the Section 5.

## 4. An hemicycle-like representation

Because of the abundance of links it is difficult to build a clear representation of the distributions of the codes and codes communities in the FLC system. We here propose a new kind of representation. If we make an analogy between the FLC network and a social network that is if we consider that the codes are individuals, where do these individuals would lie in a sort of hemicycle? First we put the rich-club of central and influent "individuals" at the center. Then we endow the set of codes with an Euclidean dissimilarity, the Czekanovski-Dice dissimilarity, which is well adapted to reveal the structure of a graph (De Fraysseix, 1999; Kuntz, 1992). This dissimilarity measures the proportion of neighbors that two vertices $v$ and $w$ don't have in common and is defined as:

$$\delta^2(v,w) = \frac{|\Gamma_v \Delta \Gamma_w|}{|\Gamma_v| + |\Gamma_w|} \qquad (2)$$

where $\Delta$ is the symmetric difference[1] (see Sec. 1.3 for the other notations). The dissimilarity between two codes is null when they have the same neighboring codes; it takes a unit value when they have no common neighbors.

With this Euclidean dissimilarity the codes can be displayed in $R^n$ ($n$ may be large and equal to the number of codes minus 1), the Euclidean distance between two codes in $R^n$ being equal to the Czekanovski-Dice dissimilarity of these two codes. Performing a principal component analysis on a distance matrix permits to display the codes on a principal plane (see the Appendix for details). The position of the $i^{th}$ code in the hemicycle that is in the half disk of unit radius, is determined by two parameters:

---

[1] The elements of $A \Delta B$ are the elements belonging to $A$ but not to $B$ and the elements belonging to $B$ but not to $A$. In other words it is the set of elements belonging to either $A$ or $B$ but not both.

- A radial component $r(j)$ which we set to be proportional to the mean distance $\bar{\delta}(j)$ of the code to the Rich-Club:

$$\bar{\delta}(i,j) = \frac{1}{|Rich-Club|} \sum_{k \in Rich-Club} \delta(k,j) \quad (3)$$

$$r(j) = 0.8 \frac{\bar{\delta}(j) - \max_k \{\bar{\delta}(k)\}}{\max_k \{\bar{\delta}(k)\} - \min_k \{\bar{\delta}(k)\}} + 1 \quad (4)$$

with $|Rich-Club|$ being the size of the rich-club.

- An angular component $\theta(j)$, which we want to be proportional to the first principal component of the principal component analysis performed of the dataset in which the effect of the mean distance to the rich-club has been suppressed, that is on the data projected on the orthogonal component to $a$ (see the Appendix) where $a$ is the vector such that the coordinates of the $j^{th}$ code along this vector is $\bar{\delta}(j)$.

Two codes may have the same mean distance to the rich-club (although the links with the club's codes are different). Therefore the addition of the angular coordinate is necessary to distinguish them on a planar representation. This hemicycle representation of the simple FLC network is given on Figure 7 where the codes are represented by their label (see Table 1). Let us note that if we do not remove the effect of the distance to the rich-club on our data then the correlation coefficient between the first principal component and this distance to the rich-club is 0.81.

Figure 7 shows two main things: a) ten codes are quite close to the rich-club without being members: they are quite influential codes, relatively central in the architecture of the FLC system; b) the codes belonging to the same stable community tend to end up in the same angular sectors. Quite influential among the codes, some belong to one of two main stable communities previously identified, others not as the monetary and financial code (with label MOF) and the defense code (with label DEF). The angular dispersion of communities (see Tab. 4) are of nearly the same amplitude, the community of the 11 codes on our representation occupying the left half of the amphitheater, the community of 12 codes the right half. This reflects an equal range of dissimilarity between codes within each community. However the most interesting feature is that we find on this representation groups of codes belonging to the same stable community, on the basis of criteria independent of those that were used to partition the system. This observation tends to corroborate the existence of these code communities that had hitherto escaped the analysis.

## 5. Discussion

At this point what have we learned? Firstly we have produced for the first time a visualization of the French network of legal codes. A pedagogical virtue of such a map is to show unequivocally the rich network of interrelationships between codes, and between major areas of law. Another is to realize how necessary it is to who wants to master this complexity to dispose of such set of navigation instruments, methodologies and analysis tools for understanding the emerging and evolutionary legal structure (Bourcier and Mazzega, 2007a).

At one level of the analysis, we find the club of the 10 codes most related to other codes. These codes (see Table 3) are, with one exception, all connected in pairs. They form a rich club that structures the inner-core of the French legal system. Foremost is the penal code, strongly connected to many other codes, as it is gathering all the measures relating to penalties and offenses, whatever the domain they apply to. There is also the civil code for its antiquity, its notoriety and fundamental dimension for the human rights that it ensures (rights of individuals). It is found to be central in the French legal system.

The Rapporteur for the Higher Commission for Codification that was also responsible for drafting the general code for local authorities (GCT), has commented some aspect of our results. First, as a kind of "blind test", we asked her to give a list of most central or influent codes, on the basis of her intuitive knowledge of the full legal system. She gave us seven codes of the rich-club. From her point of view, the most unexpected codes in the rich clubs are the general code for local authorities (GCT), the code of the environment (ENV) and the code for the public health (SAP). These three codes have in common to regulate areas covering a wide range of topics and cases, all linked to many other dimensions of the social life, politics and our living environment. For example, the code for local authorities organizes all links between the state and its decentralized services, administration in the regions, links with regional and local services, etc. Just check the table of content of these codes to be convinced of the extent and diversity of subjects they cover (to be consulted on the Legifrance web site).

If we discard these 10 most central codes from the network, we find again a form of organization which probably has so far escaped analysis. Indeed we have found two communities of codes, the identification of which is stable and robust regardless of the analysis criterion and the algorithm used for partitioning the FLC system. Looking through the list of 12 codes belonging to the first community there appears a common feature for all matters within a fuzzy "social domains" (codes of social action and families, insurance, consumers, etc.), "activities with a social character" (code of handicraft, education, tourism, etc.) and regulation of these areas (code of administrative justice, but also traffic, etc.). We could group under the term "codes for social issues" this stable community. The second community could be called "codes for territories and resources", suggesting perhaps too indirectly the idea of ownership (of land, natural resources with a geographical or spatial character): codes for housing, state-owned property, expropriation in public interest, forestry, mining, urban planning, etc.

It is quite remarkable that these same communities also reappear when completely different criteria are used to build another representation suggesting an interpretation of the network associated to the French legal system (based on a factor analysis of a dissimilarity table and representation in a hemicycle). These results confirm the validity of the interpretative approach to the main code communities found as far in the system (rich club, codes for social issues, codes for territories and resources). Facing such new representation of a whole legal system, the doctrine might raise new questions and investigation, elaborate a new perception of the French legal system, even if the names proposed here for these communities should be submitted to further reflections coupled with a more detailed analysis (returning the content of these codes). These first results also strengthen that the conception of law as structured is not only legitimate but also that it can now rely on a body of theories and tools designed to address the analysis of complex large legal systems. Such approaches have been carried out on the United States Code in, for instance, (Bommarito and Katz, 2010a) by considering the US Code as a network with hierarchical links and citation links; even if the underlying models are the same (networks) our study is performed at a different scale or granularity (coarse-grained system) and focuses on the discovering of community. We shall change the scale of study and consider interactions between citation links and hierarchical links in other (Boulet R., Mazzega P. and D. Bourcier, 2009) and future works.

## 6. Conclusion

We have produced a map of the French system of legal codes. This map may be a kind of dashboard for managers of law. It is also of interest to theorists and analysts of law, providing a visualization of some networks (and sub-networks) invisible to a reader of law, even circumspect. The novelty of this approach in law is also evidenced by the absence of a tradition in legal theory that would establish a quantitative analysis of the architecture that we reveal in the French legal system, even if the in-depth development of the structures of law are the subject of rich contemporary analysis linking these structural changes with the evolution of stakes and moving powers from the State level to the scene of the international politics (Delmas-Marty, 2007; Ost and van de Kerchove, 2002).

Through the study of a real graph taken from our legal corpus, we find a new type of network characterized by a high centralization and a high density that we call *concentrated world*, paving the way for research and study of other graphs of this type found in real applications. Indeed we suspect that such concentrated-world structures to be typical of various legal systems considered at the national scale and with a granularity at the code-level (or equivalent groups of legal texts), in particular those related to the large family of civil law.

The high number of edges in a concentrated world being a handicap for its analysis, we have introduced the idea of a "representation in a hemicycle", to better visualize and understand the structure of the graph, and to corroborate the existence of hidden code communities. In this representation some groups of codes appear, consistent with groupings obtained from the graph partitioning, hitherto unknown to lawyers. By producing a new map of the national legal system, many opportunities open to analysis, which may offer significant innovations on our perception of law at this scale, on the development of the engineering of law, on its understanding and within the "factories of law".

At this stage we have mainly opened new perspectives. In particular in Part II of this study we analyze the weighted graph whose edges are weighted by the number of citations between codes. These weights induce a metric in where two codes citing each other many times are very "close". As we shall see this analysis provides new insights on the community structure of codes we have found in the Part I. So doing, we pursue the process of defining measures of the complexity of the law (Bourcier and Mazzega, 2007b) based on an analysis of its formal structure and content.

**Acknowledgements.** We are very grateful to Mme Elisabeth Catta, *Rapporteur* for the Higher Commission for Codification, for her interest in our work and for her helpful comments. R. Boulet has benefited from a post doctoral grant of the *Institut National des Sciences de l'Univers* (CNRS, Paris). This study was funded by the RTRA *Sciences et Techniques de l'Aéronautique et de l'Espace* (http://www.fondationstae.net/ ) in Toulouse (MAELIA project - http://maelia1.wordpress.com/ ). The yEd Graph editor has been used for producing the Fig.1, Fig.4, Fig.5 and Fig. 6. Statistical properties of networks have been computed with R and the library igraph (http://www.rproject.org/ ); Fig.2 and Fig. 3 were obtained with R.

# Appendix

Let us recall some aspects of a Principal Component Analysis (PCA). If *X* is a data matrix with *n* rows and *p* columns (the $i^{th}$ row of *X* represents the coordinates in $R^p$ of the $i^{th}$ individual), a PCA gives a new coordinate system, that is a set of vectors $u_1, ..., u_p$ which is the new basis of $R^p$. The individuals have new coordinates in this basis. The vector consisting of the $j^{th}$ coordinate of the individuals is called $j^{th}$ principal component and we denote it by $c_j$. This new coordinate system is such that maximizing the variance of the data projected on an *r*-dimensional subspace is done by projecting the data on the *r* first principal coordinates. It turns out that the vectors $u_j$ are the orthonormal eigenvectors of the covariance matrix **V** = **XX'** (**X'** being the transpose of **X**). The principal coordinates $c_j$ are given by $c_j$ = **X**$u_j$ with $u_j = \sqrt{\lambda_j} z_j$ where the $z_j$ are the orthonormal eigenvectors of the scalar product matrix **W** = **X'X**. with matrix **Λ of** eigenvalues $\lambda_j$.

Proceeding in the same way with the distance matrix **D** we obtain the scalar product matrix **W** by the Torgerson formula

$$w_{ij} = -\frac{1}{2}\left(d_{i,j}^2 - d_{i,.}^2 - d_{.,j}^2 - d_{.,.}^2\right) \tag{A1}$$

where

$$d_{i,.}^2 = -\frac{1}{n}\sum_j d_{i,j}^2, \quad d_{.,j}^2 = -\frac{1}{n}\sum_i d_{i,j}^2 \text{ and } d_{.,.}^2 = -\frac{1}{n}\sum_i d_{i,.}^2 \tag{A2}$$

Once we have the matrix **W**, a principal component analysis gives us the principal coordinates. Now, let **X** be a $n \times p$ data matrix and let $a* \in (R^p)*$ be a variable represented by the vector $a \in R^p$ (Riesz representation theorem). If we want to suppress the influence of the variable $a*$ on the data matrix, we can consider the data projected on the orthogonal space of $a$ (denoted $a^\perp$). The new data matrix is **Y** = **XP** where **P** = **I** − $aa'$ is the matrix of the projection on $a^\perp$ (with $\|a\|=1$), we can then perform a principal component analysis on the new data matrix **Y**.

If we do not have **X** but the distance matrix **D** (and consequently the matrix **W**, the eigen decomposition of which is **W =QΛQ**$^{-1}$) we can reconstruct a data matrix by setting **X =Q√Λ** (which is in fact the coordinates of individuals in the principal component system). Moreover we may not have directly the vector $a \in R^p$ representing the variable $a* \in (R^p)*$ but we have the value $\tilde{c}$ of the *n* individuals for the variable $a*$. The $j^{th}$ coordinate of $a$ is the correlation coefficient between $\tilde{c}$ and the $j^{th}$ principal component $c_j$.

In practice, we obtained the angular coordinate $\theta(j)$ of the $j^{th}$ code in the hemicycle representation of the FLC network (see Sec. 4) in the following way: from the distance matrix **D** we get the scalar product matrix **W**, perform a first PCA and we recover **X**. Then we compute the vector $a$ and perform a second PCA on **XP** (where **P** is the matrix of the projection on $a^\perp$) and get a first principal component $c_1$. Finally $\theta(j)$ is given (in rd) by:

$$\theta(j) = \frac{\pi[c_1(j) - \min_k(c_1(k))]}{\max_k(c_1(k)) - \min_k(c_1(k))} \tag{A3}$$

**TABLE 1: Labels and short name of the 52 French Legal Codes retained in this study.**

| Label | Short Name | Label | Short Name |
|---|---|---|---|
| ART | Handicraft | LGA | Honor Legion |
| ASF | Social service | MIN | Mining |
| ASS | Insurance | MOF | Monetary & financial |
| AVI | Civil aviation | MPU | Public contract |
| CHA | Housing | MUT | Mutual society |
| CIV | Civil | OGJ | Admin. of justice |
| CNS | Consumer | PAT | Estate |
| COM | Trade | PCI | Civil procedure |
| DEF | Defense | PCO | Post communication |
| DMM | Mercantile marine | PEN | Criminal |
| DOE | State-owned property | PIT | Intellectual property |
| DOU | Customs | PMA | Seaports |
| DPF | Public rivers | PPE | Criminal procedure |
| EDU | Education | REC | Research |
| ELE | Elections | ROU | Traffic |
| ENV | Environment | RUR | Rural |
| EUP | Expropriation in public interest | SAP | Public Health |
| FAS | Family | SDA | Asylum |
| FOR | Forestry | SNA | National service |
| GCT | Local authorities | SPO | Sport |
| GPP | Property legal person | SSC | Social Security |
| ICI | Film industry | TMA | Marine employment |
| INM | Monetary media | TOU | Tourism |
| JUA | Administrative court | TRA | Employment |
| JUF | Financial court | URB | Urbanism |
| JUM | Military court | VOR | Road system |

**TABLE 2:** Structural indices for the FLC network, the associated random graph and some other networks already studied. The number of vertices is denoted by n, the number of edges is denoted by m, d denotes the density and k the mean degree. The mean of shortest paths is denoted by $\bar{l}$, and the characteristic path length by L. D denotes the diameter, C1 and C2 the clustering coefficients. Degree centralization, betweenness centralization and closeness centralization are denoted by CD, CB and CP respectively.

|  | FLC network | G(n,d) simulated | Math. network | P2P network | Medieval network |
|---|---|---|---|---|---|
| $n$ | 51 | 51 | $2.1\ 10^5$ | $6.2\ 10^6$ | 615 |
| $m$ | 531 | 531 | $4.6\ 10^5$ | $1.6\ 10^8$ | 4193 |
| $d$ | 0.416 | 0.416 | $2.1\ 10^{-5}$ | $8.2\ 10^{-6}$ | 0.022 |
| $k$ | 20.8 | 20.8 | 4.4 | 51.3 | 13.64 |
| $\bar{l}$ | 1.595 | 1.584 | 7.73 | 4 | 3.9 |
| $L$ | 1.569 | 1.553 | - | - | 3.71 |
| $D$ | 3.0 | 2.06 | 27 | 10 | 10 |
| $C_1$ | 0.694 | 0.416 | 0.72 | 0.13 | 0.78 |
| $C_2$ | 0.601 | 0.415 | - | 0.07 | 0.46 |
| $C_D$ | 0.441 | 0.163 | - | - | 0.105 |
| $C_B$ | 0.047 | 0.006 | - | - | 0.061 |
| $C_P$ | 0.456 | 0.141 | - | - | 0.252 |

**TABLE 3: List of the ten codes forming the rich-club. These codes also have the highest degrees (see text).**

| Label | Short Name | Degree |
|-------|------------|--------|
| PEN | Criminal | 42 |
| GCT | Local authorities | 40 |
| SAP | Public health | 40 |
| PPE | Criminal procedure | 39 |
| CIV | Civil | 38 |
| RUR | Rural | 37 |
| COM | Trade | 36 |
| ENV | Environment | 35 |
| SSC | Social security | 35 |
| TRA | Employment | 35 |

**TABLE 4: The two main stable communities (excluding the rich-club – see text).**

| A stable community with 12 codes | | A stable community with 11 codes | |
|---|---|---|---|
| *Label* | *Short Name* | *Label* | *Short Name* |
| ART | Handicraft | CHA | Housing |
| ASF | Social service | DOE | State-owned property |
| ASS | Insurance | DPF | Public Rivers |
| CNS | Consumer | EUP | Expropriation in public interest |
| EDU | Education | FOR | Forestry |
| JUA | Administrative Court | GPP | Property legal person |
| JUF | Financial Court | MIN | Mining |
| MUT | Mutual society | PAT | Estate |
| REC | Research | PMA | Seaports |
| ROU | Traffic | URB | Urbanism |
| SPO | Sport | VOR | Road system |
| TOU | Tourism | - | - |

**FIGURE 1: Representation of the network of the French Legal Codes, with the labels of the codes (see Table 1).**

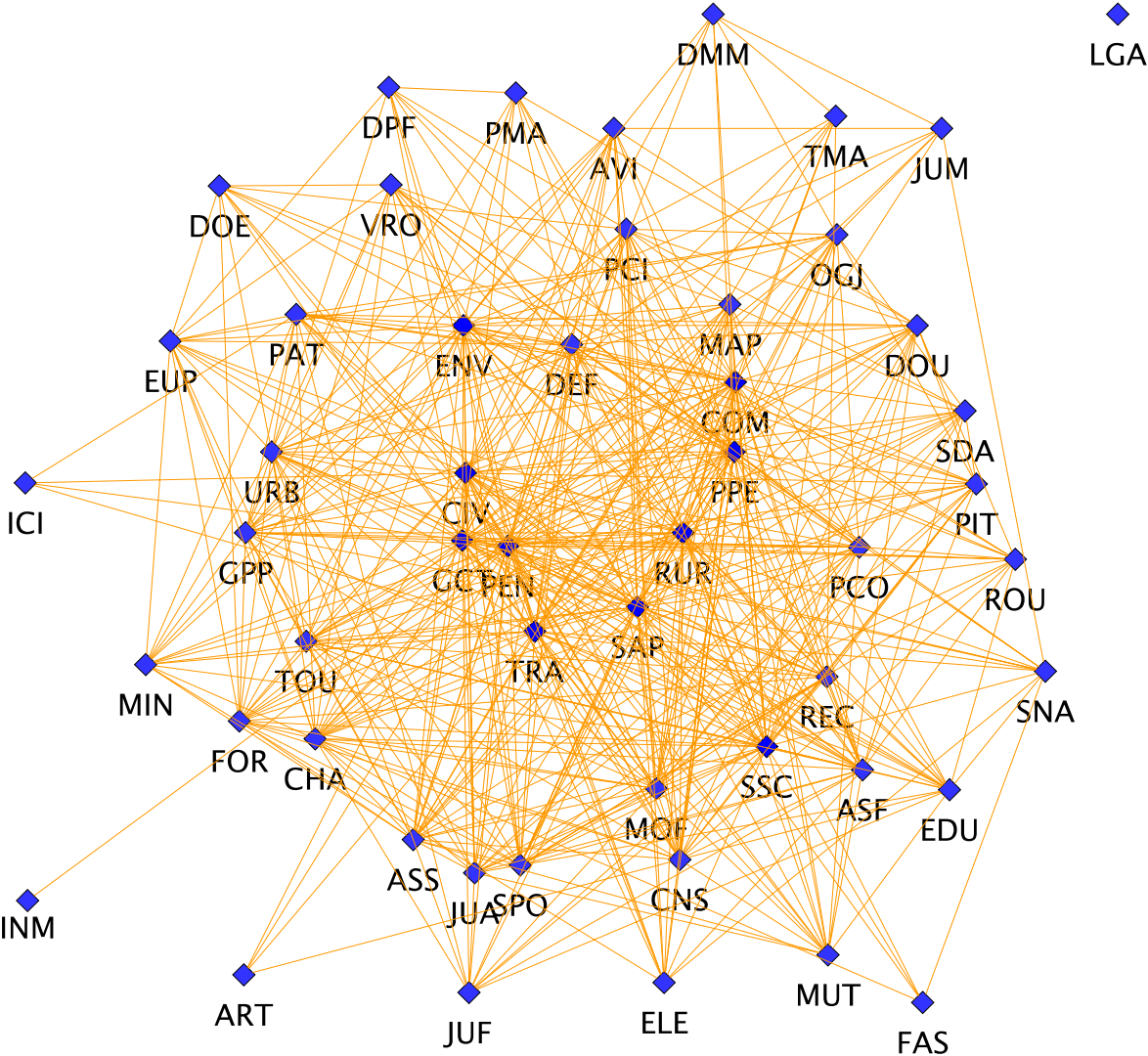

**FIGURE 2: Density and diameter of the subgraph induced by a percentage of highest degree vertices (log scales). In the FLC system, the rich-club, determined by a drop of the density and a concomitant increase of the diameter, forms a central community with a very high density and low diameter.**

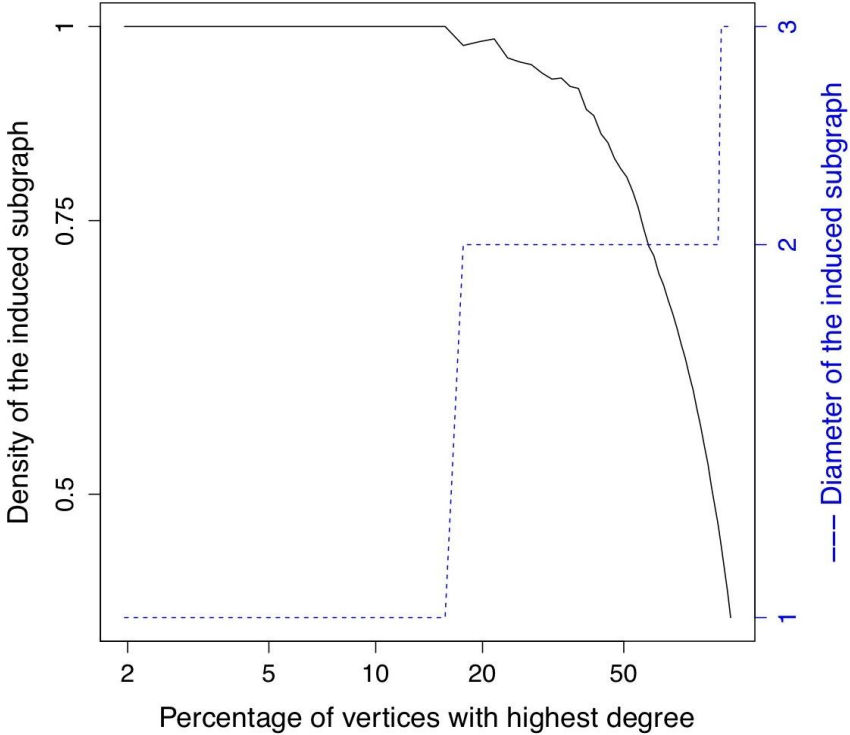

**FIGURE 3: Betweenness centrality measure (top) and closeness centrality measure (bottom) of vertices sorted by decreasing order of degree of the codes (vertices) of the French Legal System.**

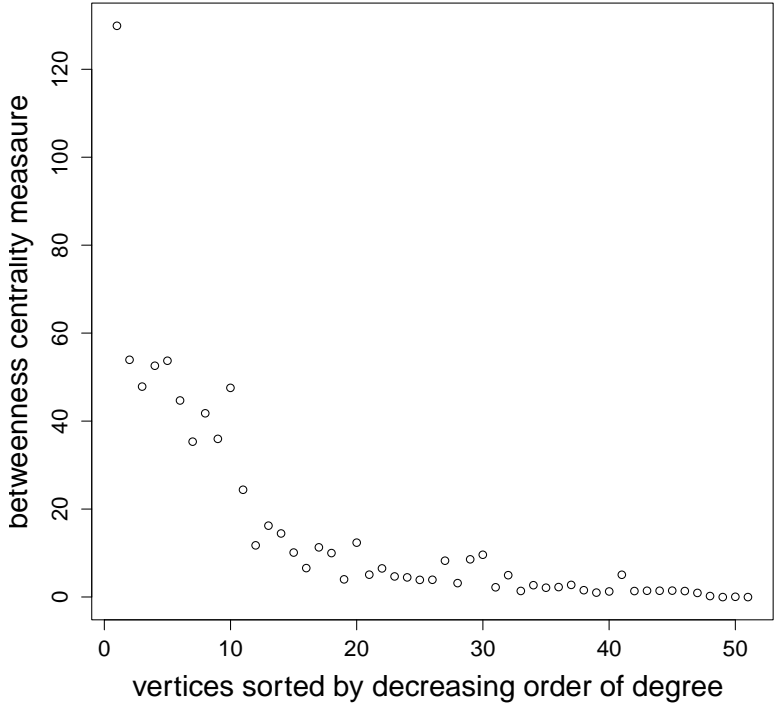

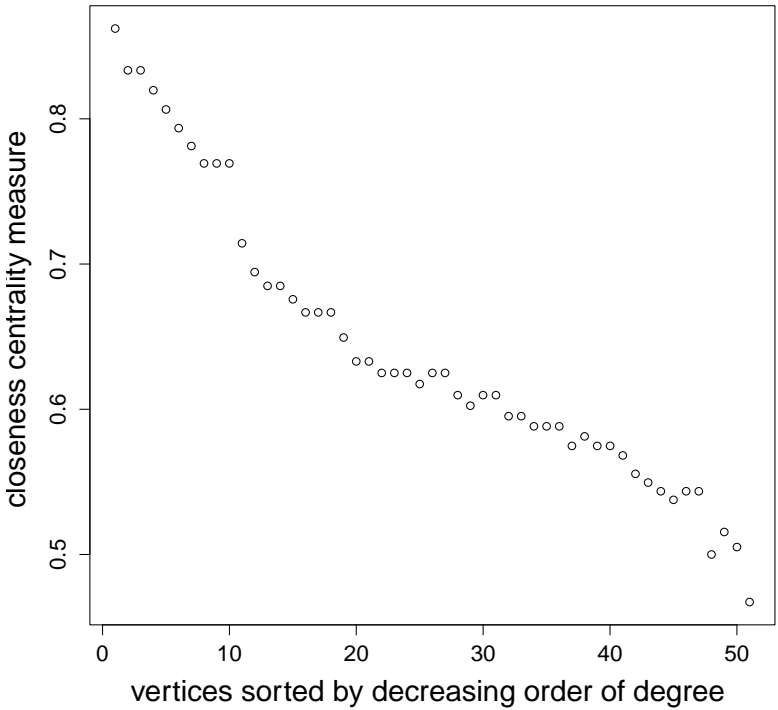

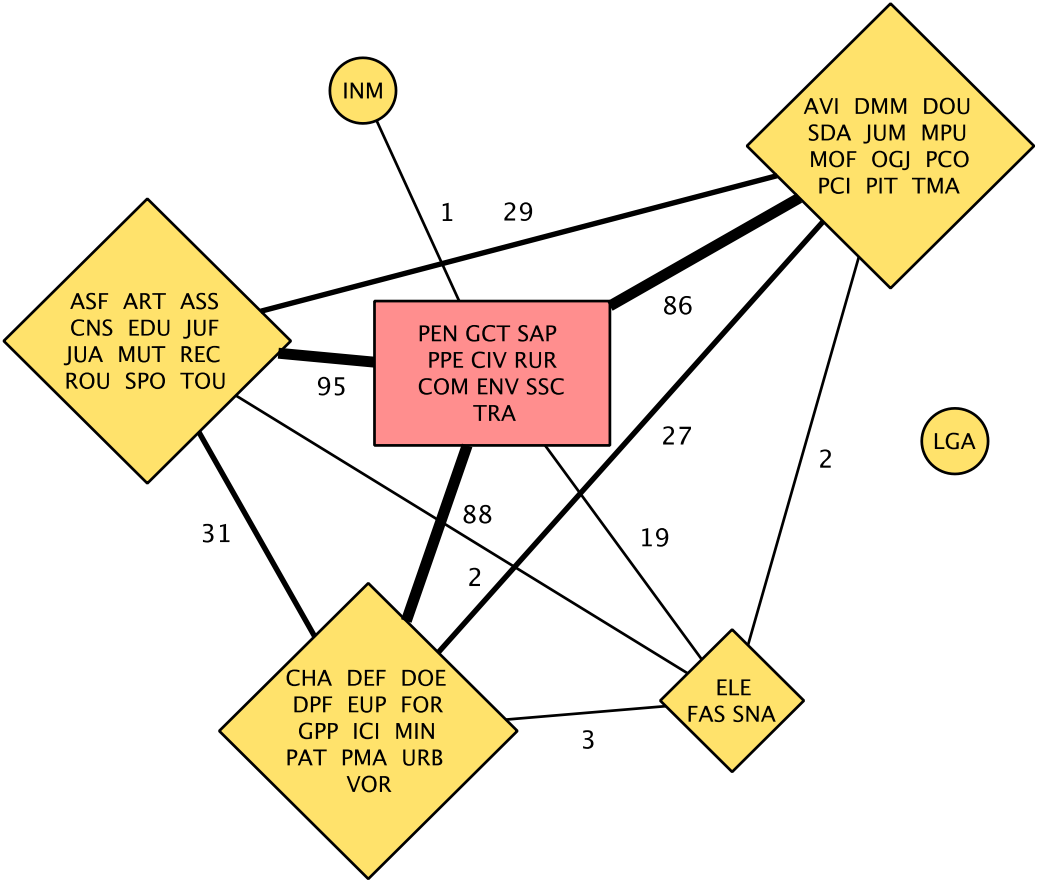

FIGURE 4: Modular representation of the network associated to the FLC system (see Fig. 1), as obtained with the fast-greedy partitioning algorithm (see text). Labels of the legal codes are given in Table 1. The number of edges between two communities are represented by a weighted edge between these two communities.

**FIGURE 5:** Same as Figure 4 but with the spectral partitioning algorithm (see text).

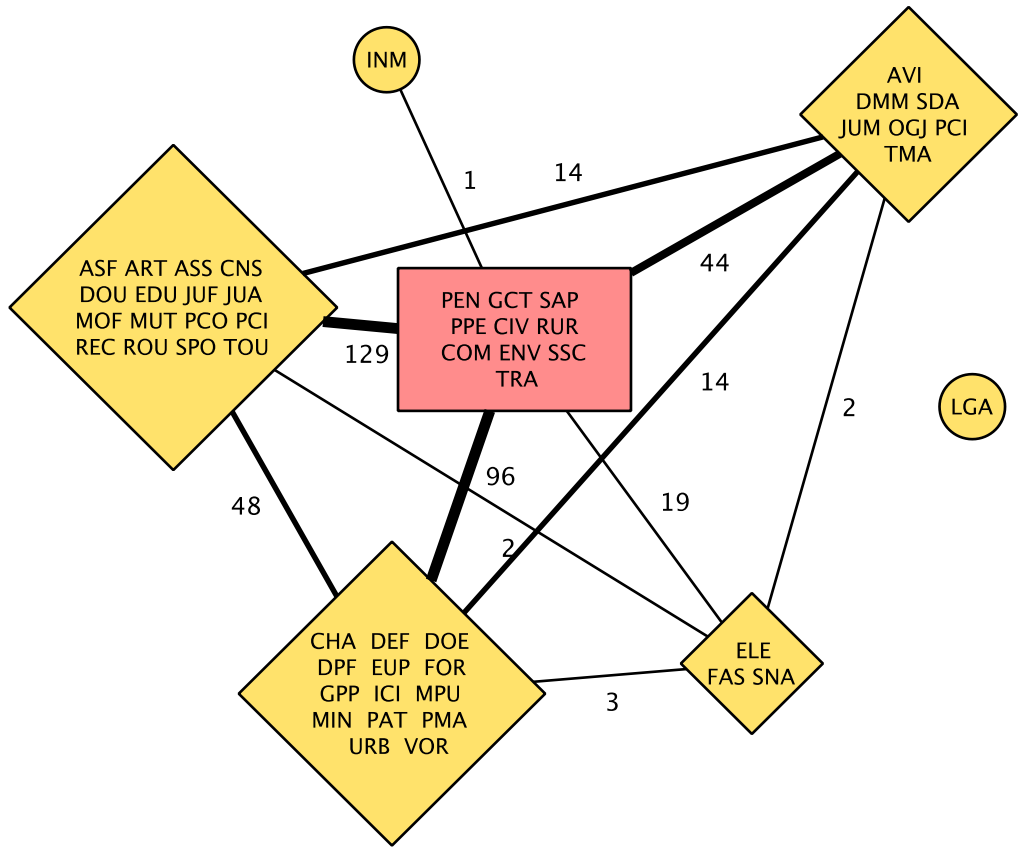

FIGURE 6: Same as Figure 4 but with the walk-trap partitioning algorithm (see text).

**FIGURE 7:** The hemicyle-like representation of the weighted FLC network. The two stable groups (Table 4) are highlighted: the one with twelve codes is in pink (and in italic), the one with eleven codes is in blue (and in bold). The center of the hemicycle represents the rich-club.

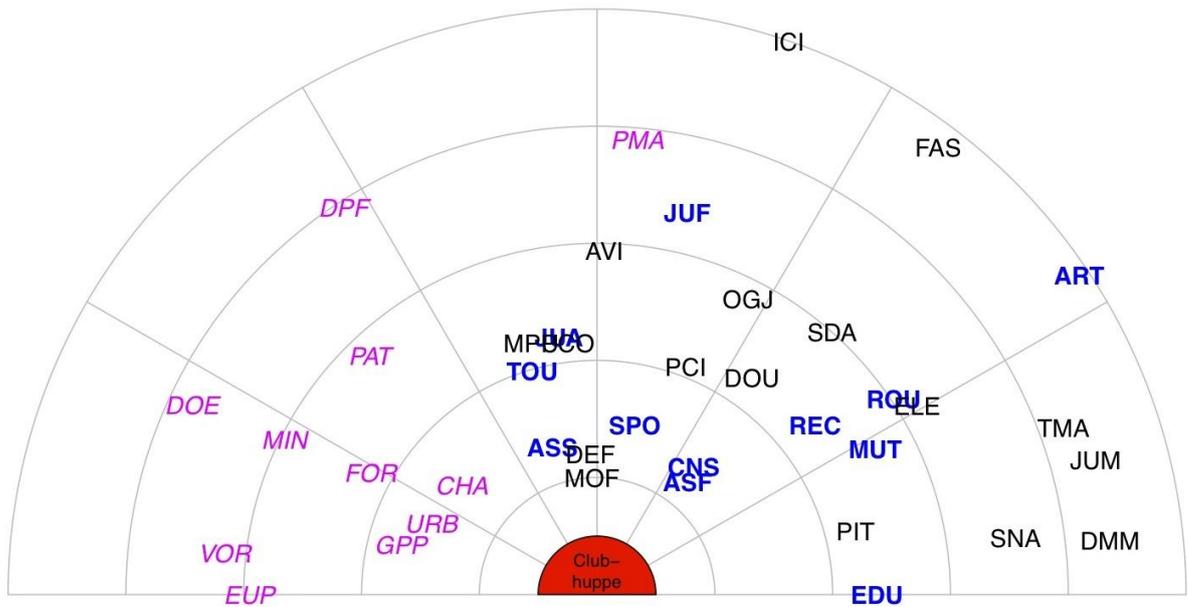